\documentclass[useAMS,usenatbib]{mn2e}
\usepackage{graphicx,amsmath}
\usepackage{mathrsfs}
\makeatletter

\newcommand{\Rmnum}[1]{\expandafter\@slowromancap\romannumeral #1@}
\newcommand{\gsim}{\mbox{$\stackrel{>}{_{\sim}}$}}
\makeatother

\title[White dwarf with Strong Poloidal Magnetic Field]{Mass-radius relation of strongly magnetized white dwarfs: nearly independent of Landau quantization}
\author[P. Bera and D. Bhattacharya]{Prasanta Bera\thanks{E-mail:pbera@iucaa.ernet.in}, Dipankar Bhattacharya\thanks{E-mail:dipankar@iucaa.ernet.in}\\
Inter University Centre for Astronomy and Astrophysics, Post Bag 4, Pune 411007, India.}
\begin{document}


\pagerange{\pageref{firstpage}--\pageref{lastpage}} \pubyear{0000}

\maketitle

\label{firstpage}

\begin{abstract}
We compute static equilibria of white dwarf stars containing strong poloidal magnetic field, and present the modification of white dwarf mass-radius relation caused by the magnetic field.  We find that a maximum white dwarf mass  of about $1.9$~$M\odot$ may be supported if the  interior field is as strong as approximately $10^{10}$ T. This mass is over 30 per cent larger than the traditional Chandrasekhar Limit.  The equation of state of electron degenerate matter can be strongly modified due to Landau quantization at such high magnetic fields.  We find, however, that this does not significantly affect the structure of the white dwarf.
\end{abstract}


\section{Introduction}
White dwarfs are the stellar remnants of low and intermediate mass stars (with mass below $10M_{\odot}$) and are composed primarily of electron degenerate matter. They are highly dense, typically of the order of a solar mass squeezed within a volume comparable to that of the earth.

In the classical limit (i.e. the kinetic energy of electrons being small compared to rest energy) the radius of a white dwarf is inversely proportional to the cube root of its mass. But in the relativistic limit the electron degenerate white dwarf mass has a maximum value of $1.44M_{\odot}$ which is known as the Chandrasekhar mass ($M_{\rm Ch}$) limit.

As the mass of an accreting white dwarf in a binary system approaches the limiting mass, the star would undergo a rapid contraction. Elevated temperature during this collapse can trigger a runaway thermonuclear reaction resulting in a supernova explosion of Type Ia (SNIa) (for details of SNIa ignition process see, for example, \cite{martin_06,mazzali_07}). The standard conditions that lead to the SNIa allow them to be used as standard candles in cosmology, and has led to the discovery of the accelerating expansion of the Universe \citep{riess_98}.

A small subset of SNIa is now being found to be more luminous than usual (e.g. 2003fg, 2006gz, 2009dc). It has been suggested that the progenitor of such over-luminous SNIa may be super-Chandrashekhar mass white dwarfs ($M \gsim 2M_{\odot}$;\cite{howel06,hicken_07,yamanaka_09}). Such a high mass of a white dwarf may derive from its rapid rotation, or a transient high mass configuration may be formed in the merger of two heavy white dwarfs \citep{moll_13}. It has also been suggested that strong magnetic field may be responsible for supporting this additional mass \citep{das_m12, das_m13, das_m14}.  However, such investigations have so far involved rather unrealistic field configurations, such as a uniform field \citep{das_m13} that requires sources outside the star \citep{Coelh13,nitya13}, or a spherically symmetric field distribution \citep{das_m14} which is not fully consistent with Maxwell's equations.  It has also been argued that the modification of electron degenerate equation of state due to Landau quantization may have a profound effect on the white dwarf structure, and could raise the mass limit above $2$~$M_{\odot}$, even if the effect of magnetic pressure is ignored \citep{das_m13}.  The modification of mass-radius relation of white dwarfs due to Landau-quantized equation of state has been computed for low-strength, uniform fields by \cite{suh_m00}.

In this paper we compute the structure of strongly magnetic white dwarfs, considering the internal magnetic field to be axisymmetric and strongly poloidal.  We self-consistently account for the Lorentz force, with current density bounded within the star.  We also include the effect of Landau quantization on the equation of state of electron degenerate matter and estimate its impact on the structure of the magnetic white dwarf.  We compute a sequence of configurations for several different internal field strengths and derive the mass-radius relation for each case.  Work of similar nature has been previously attempted by \cite{Ostiker_Hartwick68}, \cite{tomim05} and \cite{lande09}.  However these earlier works did not include the effect of Landau quantization, and also did not compute full structure sequences required to derive the mass-radius relation.
 
An outline of this paper is as follows. We present the basic equations of stellar structure and our numerical formulation in Section~\ref{numerical_method}. The characteristics of the configurations obtained using this formulation are described in Section~\ref{results}. We conclude by discussing our results in Section~\ref{conclusion}.
 
\section{Formulation and Numerical method} \label{numerical_method}

In this section, we briefly describe the self-consistent field formalism \citep{hachi86, tomim05, lande09} employed to obtain equilibrium magnetic white dwarf configurations from the basic stellar structure equations. Here, we use the spherical polar ($r$,$\theta$,$\phi$) and cylindrical polar ($\xi$,$\phi$,$z$) coordinates with their origin coinciding with stellar centre such that, $\xi=r\sin\theta$ and $z=r\cos\theta$.

\subsection{Basic Assumptions}
We assume the white dwarf configuration to satisfy the following conditions.
\begin{enumerate}
\item 
The white dwarf is cold throughout i.e. the temperature is well below the Fermi temperature ($T \ll T_{\rm F}$). So the electrons occupy states only up to the Fermi energy ($E_{\rm F}$).
\item The system is in a stationary state. ($\frac{\partial}{\partial t}\rightarrow 0$)
\item The system is axisymmetric, i.e.$\frac{\partial}{\partial \phi}\rightarrow 0$.
\item The system is non-rotating.
\item The source of the magnetic field (i.e. the current distribution) is confined within the white dwarf.
\item Conductivity of the stellar medium is infinite (i.e. ideal MHD approximation)
\item The equation of state in the presence of the magnetic field is calculated using the Landau level quantization of energy levels \citep{lai_s91}.
\end{enumerate}

\subsection{Stellar structure equations}
In a stable, static white dwarf in the presence of a magnetic field, force due to the electron degeneracy pressure is balanced by the gravitational force and the Lorentz force
\begin{equation}
\frac{1}{\rho}\mathbf{\nabla} P=-\mathbf{\nabla} \Phi_{\rm g}+\frac{1}{\rho}\left( \mathit{\mathbf{ j}}\boldsymbol\times\mathit{\mathbf{B}}\right) 
\label{hydro_eui}
\end{equation}
\\where $P$, $\rho$, $\Phi_{\rm g}$, $\mathit{\mathbf{j}}$ and $\mathit{\mathbf{B}}$ are pressure, mass density, gravitational potential, current density and  magnetic field respectively.

The gravitational potential can be obtained from the mass distribution using the Poisson's equation,
\begin{equation}
 \nabla^2\Phi_{\rm g} = 4 \mathrm{\pi} G\rho
 \label{poisson}
\end{equation}
\\where $G$ is the Gravitational constant.

In the ideal MHD limit (conductivity $\sigma\rightarrow \infty$) the magnetic field $\mathbf{B}$ satisfies the following form of Maxwell equations
\begin{equation}
 \mathbf{\nabla}.\mathbf{B}=0
\label{max_1}
\end{equation}
and \begin{equation}
 \mathbf{\nabla}\boldsymbol\times\mathbf{B}=\mu_0\mathbf{j}
\label{max_2}
\end{equation}
\\where $\mu_0$ is free space permeability.

To obtain a stable static stellar structure, Eq.~($\ref{hydro_eui}$)-($\ref{max_2}$) are to be solved self-consistently. The equation of state ($P=P(\rho)$) can be obtained from the Fermi degenerate condition. The magnetic field ($\mathbf B$) and the corresponding current density ($\mathbf j$) can be chosen to have a functional form obeying the axisymmetry condition.

\subsection{Equation of State}

A white dwarf is a dense configuration supported by the electron degeneracy pressure against gravitational collapse. The interior temperature ($T$) is very small compared to the Fermi temperature ($T_{\rm F}$), and thus the electrons occupy energy states up to $E_{\rm F}$. In the absence of a magnetic field, the electron number density $n_{\rm e}$ is obtained by integrating the phase space volume $\frac{2}{h^3}\int d^3p$. Here the factor 2 is due to the spin degeneracy of electrons and $h^3$ is the phase volume of each state. 

\begin{table*}
\begin{tabular}{c|c|c}
\hline
 & \bf B=0 & \bf B$\neq$0 \\ \hline 
Phase space integral & $\frac{2}{h^3}\int d^3p=\frac{1}{\pi^2\lambda_{\rm e}^3}\int\left(\frac{p}{m_{\rm e}c}\right)^2d\left(\frac{p}{m_{\rm e}c}\right)$  & $\sum_{\nu}\frac{2eB}{h^2}g_{\nu}\int dp_z=\frac{2\beta}{(2\pi)^2\lambda_{\rm e}^3} \sum_{\nu}g_{\nu}\int d\left(\frac{p_z}{m_{\rm e}c}\right)$ \\ 
Mass density & $\rho=\mu_{\rm e} m_{\rm H}\frac{1}{3\pi^2\lambda_{\rm e}^3}x_{\rm F}^3$  & $\rho=\mu_{\rm e} m_{\rm H}\frac{2\beta}{(2\pi)^2\lambda_{\rm }e^3} \sum_{\nu=0}^{\nu_m}g_{\nu}x_{\rm e}(\nu)$ \\ 
Pressure & $P=\frac{\pi m^4c^5}{3h^3}\left[ x_{\rm F}(2x_{\rm F}^2-3)\sqrt{1+x_{\rm F}^2}-3\sinh^{-1}x_{\rm F}\right]$  & $P=\frac{2\beta m_{\rm e}c^2}{(2\pi)^2\lambda_{\rm e}^3}\sum_{\nu=0}^{\nu_m}g_{\nu}(1+2\nu\beta)\eta\left(\frac{x_{\rm e}(\nu)}{1+2\nu\beta}\right)$ \\
Pressure Gradient & $\boldsymbol\nabla P =\frac{\rho}{\mu_{\rm e} m_{\rm H}} \boldsymbol\nabla E_{\rm F} $  & $\boldsymbol\nabla P = \frac{\rho}{\mu_{\rm e} m_{\rm H}} \boldsymbol\nabla E_{\rm F} +\left( \frac{\partial P}{\partial \beta}\right)_{E_{\rm F}} \boldsymbol\nabla\beta$ \\ \hline
\end{tabular}
\caption{Description of the non-magnetic and magnetic equations of state.  Here $\lambda_{\rm e}=\frac{\hbar}{m_{\rm e}c}$, $x_{\rm e}(\nu)=\frac{p_{z_{\rm F}}}{m_{\rm e}c}$, $\mu_{\rm e}$: the mean molecular weight per electron, $m_{\rm H}$: mass of hydrogen atom, $\eta(z)=\frac{1}{2}z\sqrt{1+z^2}-\frac{1}{2}\ln(z+\sqrt{1+z^2})$ and $\left( \frac{\partial P}{\partial\beta}\right)_{\epsilon_{\rm F}}=\frac{m_{\rm e}c^2}{(2\pi)^2\lambda_{\rm e}^3}\sum_{\nu=0}^{\nu_m}g_{\nu}\left[\epsilon_{\rm F}x_{\rm e}(\nu)-(1+4\nu\beta)\ln \frac{x_{\rm e}(\nu)+\epsilon_{\rm F}}{\sqrt{1+2\nu\beta}} \right] $.}\label{eos}
\end{table*}

But in the presence of a magnetic field $\mathbf{B}$ the component of momentum perpendicular to $\mathbf{B}$  becomes quantised, an effect known as Landau quantization. The solution of the relativistic Dirac equation of an electron of mass $m_{\rm e}$ in the presence of magnetic field $\mathbf{B}$ along z-axis gives, for $z$-momentum $p_z$, the energy eigenvalue of $\nu$-th quantised level as
\begin{equation}\label{E_ll}
 E_{\nu,p_z}=\left[ c^2p_z^2+m_{\rm e}^2c^4\left(1+\nu\frac{2B}{B_c} \right)\right]^{\frac{1}{2}},
\end{equation}
where $B$ is the magnitude of magnetic field $\mathbf B$, $B_c=\frac{m_{\rm e}^2c^2}{\hbar e}=4.414\times10^9$T, $c$ is speed of light, $h$ is the Planck's constant, $\hbar=h/2\pi$, $\nu=(n_L+\frac{1}{2}+\sigma)$, $n_L$=0,1,2....~and  $\sigma=\pm\frac{1}{2}$ \citep{canuto_chiu68}.

As momentum is a real quantity, $p_z^2$ must be non-negative. Therefore, Eq.~($\ref{E_ll}$) implies that the electron can occupy the $\nu$-th energy level so long as $\nu \leq \nu_m=\text{integer}\left( \frac{\epsilon_{\rm F}^2-1}{2\beta}\right)$, where  $\epsilon_{\rm F}=\frac{E_{\rm F}}{m_{\rm e}c^2}$ and $\beta=\frac{B}{B_c}$.

In classical description, a charged particle in a magnetic field moves in a circular orbit about a guiding centre, which has the freedom to be located anywhere in the orbital plane. The quantum mechanical  analogue of this guiding centre freedom, restricted within a large but finite projected \textit{xy}-plane, gives a degeneracy factor $\frac{eB}{h}g_{\nu}$ per unit area to a state of energy $E_{\nu,p_z}$ and \textit{z}-momentum $p_z$. The spin degeneracy factor $g_{\nu}$ is $1$ for $n_{\rm L}=0$, and  $2$  for $n_{\rm L}\geq 1$ \citep{landau_lifshitz74}.
Therefore the number of states (per unit spatial volume) in an interval $\Delta p_z$ of the \textit{z}-component of momentum, for a given $\nu$, is $g_{\nu}\frac{eB}{h}\frac{\Delta p_z}{h}$.  Using this phase space structure mass density and degeneracy pressure can be calculated as listed in Table~$\ref{eos}$ \citep{hamada_sol61,salpeter61,lai_s91}.

\subsection{Current Density for Mixed Field Configuration}
From the assumption of axisymmetry and the condition of stationarity, the magnetic field ($\mathbf B$) and current density ($\mathbf j$) can be expressed in terms of a flux function $u$ in cylindrical coordinate ($\xi,\phi,z$) as follows:
\begin{align}
 \mathbf B &=\frac{1}{\xi}\left( \boldsymbol\nabla u\boldsymbol\times \boldsymbol{\hat\phi}\right)+B_{\phi}\boldsymbol{\hat\phi}. \label{B}\\
 \mu_0\mathbf j &=\frac{1}{\xi}\left[ \boldsymbol\nabla(\xi B_{\phi})\boldsymbol\times\boldsymbol{\hat\phi}\right] -\frac{1}{\xi}\nabla_{\ast}u\boldsymbol{\hat\phi}
\end{align}
where $\boldsymbol{\hat\phi}$ is the unit vector perpendicular to the $\xi-z$ plane, $\nabla_{\ast}=\frac{1}{\xi}\left[ \frac{\partial^2}{\partial \xi^2}-\frac{1}{\xi}\frac{\partial }{\partial \xi}+\frac{\partial^2}{\partial z^2}\right]$,
 $u=\xi A_{\phi}=r\sin \theta A_{\phi}$, and $A_{\phi}$ is the $\phi$-component of magnetic vector potential $\mathbf A=(A_{\xi},A_{\phi},A_z)$. The $\phi$-component of the magnetic vector potential 
satisfies the following partial elliptic differential equation \citep{tomim05}:
\begin{align}
 \nabla^2 (A_{\phi}\sin\phi) &= -\mu_0j_{\phi}\sin\phi 
\end{align}

For an axisymmetric stable structure the Lorentz force has to be purely poloidal. For mixed field condition  $\xi B_{\phi}$ depends only on the flux function $u$ \citep{lande09}:
\begin{align}
 \xi B_{\phi}=f(u)
\end{align}
The Lorentz force then becomes
\begin{align}
 \mathbf j\boldsymbol\times\mathbf B =-\frac{1}{\mu_0}\frac{1}{\xi^2}\left[ \nabla_{\ast}u+f(u)\frac{df}{du}\right]\boldsymbol\nabla u .
\end{align}

Introducing $\boldsymbol\nabla P $ from Table~$\ref{eos}$ into Eq.~($\ref{hydro_eui}$) we find 
\begin{align}\label{Grad_eular_eq}
 \frac{1}{\mu_{\rm e} m_{\rm H}}\boldsymbol\nabla E_{\rm F} + \boldsymbol\nabla \Phi_{\rm g} =\frac{1}{\rho}\left[\mathbf j\boldsymbol\times\mathbf B - \left(\frac{\partial P}{\partial B}\right)_{E_{\rm F}} \boldsymbol\nabla B \right] .
\end{align}
The left hand side of the above equation is the sum of the gradient of two scalar functions, which demands that the right hand side must be expressible as the gradient of some scalar function $\mathcal M$. In the absence of the term $\frac{1}{\rho}\left(\frac{\partial P}{\partial B}\right)_{E_{\rm F}}$, the current density ($\mathbf j$) can be expressed \citep{lande09} as follows:
\begin{align}
 \mu_0\mathbf j = \mathbf B\frac{df(u)}{du}+\mu_0\xi\rho\frac{d\mathcal M_0(u)}{du}\boldsymbol{\hat\phi}
\label{mu_0j_},
\end{align}
here $\mathcal M=\mathcal M_0(u)$. To consider the effect of magnetic field dependent degeneracy pressure, $\mathcal M$ can be approximated as a sum of two scalar functions $\mathcal M_0(u)$ and $\mathcal M_1$ where $\mathcal M_1$ is defined as follows:
\begin{align}
 -\frac{1}{\rho}\left(\frac{\partial P}{\partial B}\right)_{E_{\rm F}}\boldsymbol\nabla B=\boldsymbol\nabla \mathcal M_1.
 \label{sp_m1}
\end{align}
 The gradient of $\mathcal M_1$ can be expressed as the sum of two components, one along $\boldsymbol\nabla u$ and other normal to $\boldsymbol\nabla u$. The current density ($\mathbf j$) can then be written as:
\begin{align}
 \mu_0\mathbf j &=\mathbf B\frac{df}{du} \\ \nonumber
 +&\mu_0\xi\rho\left\lbrace \frac{d\mathcal M_0}{du}+\frac{\boldsymbol\nabla \mathcal M_1\cdot\boldsymbol\nabla u}{|\boldsymbol\nabla u|^2}+\frac{1}{\rho}\left(\frac{\partial P}{\partial B}\right)_{E_{\rm F}} \frac{\boldsymbol\nabla B\cdot\boldsymbol\nabla u}{|\boldsymbol\nabla u|^2} \right\rbrace \boldsymbol{\hat\phi}
\label{mu0_j_gen}
\end{align}
where the term $\mathcal M_1$ can be obtained from Eq.~($\ref{sp_m1}$) taking the component perpendicular to $\boldsymbol\nabla u$ 
\begin{equation}
 \boldsymbol\nabla \mathcal M_1\cdot(\boldsymbol{\hat\phi}\boldsymbol{\times}\boldsymbol\nabla u)=-\frac{1}{\rho}\left( \frac{\partial P}{\partial B}\right)_{E_{\rm F}}\boldsymbol\nabla B\cdot(\boldsymbol{\hat\phi}\boldsymbol{\times}\boldsymbol\nabla u)
\label{m1}
\end{equation}

\subsection{Boundary Conditions and Integral Formalism}
Considering the boundary conditions for gravitational potential and $\phi$-component of magnetic vector potential to be the following:
\begin{align}
 \Phi_{\rm g}\sim\mathcal{O}\left( \frac{1}{r}\right) \hspace{5mm}as\hspace{2mm} r\rightarrow\infty;\\
 A_{\phi}\sim\mathcal{O}\left( \frac{1}{r}\right) \hspace{5mm}as\hspace{2mm} r\rightarrow\infty;
\end{align}
the Green's function solution of these variables at $\mathbf r$ can be expressed as the volume integral over the source points $\mathbf r'$
\begin{align}
  \Phi_{\rm g}(\mathbf{r})&=-G\int\frac{\rho(\mathbf r')}{|\mathbf r-\mathbf r'|}d^3\mathbf r',
\label{int_phi_g} \\
 A_{\phi}(\mathbf r)\sin\phi &=\frac{\mu_0}{4\pi}\int\frac{j_{\phi}(\mathbf r')\sin\phi'}{|\mathbf r-\mathbf r'|}d^3\mathbf r'.
\label{int_Aphi}
\end{align}

Eq.~($\ref{Grad_eular_eq}$) can be integrated following the Hachisu Self-Consistent-Field technique \citep{hachi86} to obtain
\begin{align}\label{intro_c}
 \dfrac{1}{\mu_{\rm e} m_{\rm H}} E_{\rm F} +\Phi_{\rm g} &=\mathcal M+C,
\end{align}
where $C$ is the integration constant \citep{tomim05}.

The Fermi energy($E_{\rm F}$) and the potentials $\Phi_{\rm g}$ and $A_{\phi}$ can thus be obtained at every point using the above integral formalism. Here $\Phi_{\rm g}$ can be calculated at any point if the mass distribution is known and $\mathcal M$ can be obtained if the functional form of $\mathcal M$ is known. Now when $\frac{1}{\rho}\frac{\partial P}{\partial B}$ is negligible, $\mathcal M\rightarrow \mathcal M_0$ and $\mathcal M_0$ is a function of $u (=\xi A_{\phi})$. So, in this limit the magnetic field distribution can be derived from the assumed form of $\mathcal M_0(u)$ and $f(u)$ using  Eq.~($\ref{int_Aphi}$) and Eq.~($\ref{B}$). After obtaining a stable configuration assuming $\mathcal M=\mathcal M_0$, Eq.~($\ref{m1}$) can be used to get $\mathcal M_1$ and $\mathcal M$ ($=\mathcal M_0+\mathcal M_1$) for the general solution. The boundary value of $M_1$ may be assigned from the consideration that in the stellar core electrons can access high Landau levels, minimizing the effect of quantization. We choose the value of $\mathcal M_1$ at the core to be zero.

The quantities $\mathcal M_0 (u)$ and $f(u)$ are arbitrary functions of $u$; the stellar configuration is dependent on their functional form. To study the effect of Landau quantization on the static stellar structure for a strong magnetic field, we consider here simple forms of $\mathcal M_0 (u)$ and $f(u)$ as follows:
\begin{align}
 \mathcal M_0 (u)& \propto u=\alpha_0 u ~~~;\text{$\alpha_0$ is a constant}\label{func_Mu}\\
f(u)&=0.
\label{func_fu}
\end{align}
Which would give a purely poloidal ($B_\phi=0$) magnetic field in the configuration.

The distribution of mass and magnetic field can then be obtained self-consistently, starting from an arbitrary initial mass distribution, by using Eq.~($\ref{intro_c}$) iteratively. To carry out the numerical computations, the physical quantities are transformed into dimensionless form using scaling parameters derived from maximum density $\rho_{\rm max}$, equatorial radius $R_{\rm eq}$ and core magnetic field $B_{\rm core}$. In general $\rho_{\rm max}$ is related to maximum Fermi energy $E_{F_{\rm max}}$ and $B_{\rm core}$.  At any step of the iteration, the value of the integration constant $C$ is adjusted using the boundary conditions: (i) $E_{\rm F}=E_{F_{\rm max}}$ at the core, (ii) $E_F=0$ at the surface and (iii) $B=B_{\rm core}$ at $E_{\rm F}=E_{F_{\rm max}}$.

 The iteration would be terminated when the relative change of certain parameters ($R_{\rm eq},\alpha_0$ and $C$) falls below a specified small number ($\epsilon$). We have used a value of $\epsilon=10^{-6}$. 
 
 A configuration in static equilibrium must satisfy the stellar virial equation \citep{shapi83}: 

\begin{equation}
 3\Pi+W+\mathscr{M}=0
\label{virial}
\end{equation}

\begin{align*}
\text{here,}\hspace{5mm}\Pi:&\text{ contribution of internal energy}=\int PdV;\\W:&\text{ gravitational potential energy}=\int\rho \Phi_{\rm g}dV;\hspace{2mm} \\  \mathscr{M}:&\text{ magnetic energy}=\int\frac{B^2}{2\mu_0} dV;
\end{align*}
and $V$ is space volume.

\begin{figure}
\centering
\includegraphics[width=0.45\textwidth]{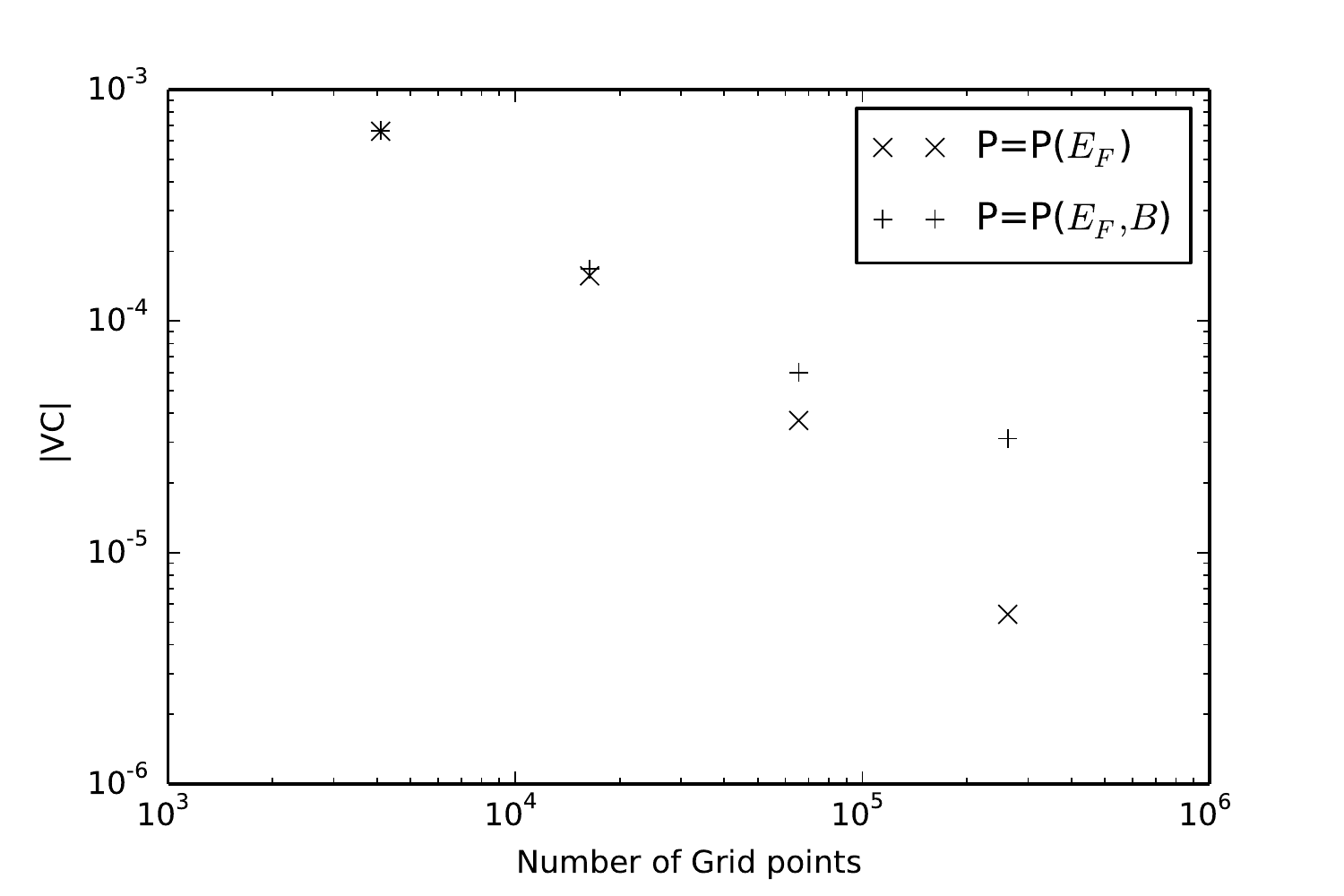}
\caption{Test of convergence of the numerical solution of the white dwarf structure. The cases of equation of state without ($\times$) and with (\text+) Landau quantization are both shown. With the increase in the number of grid points, better convergence is achieved.}
\label{VC}
\end{figure}

To check the convergence of our solutions, we tested the virial condition (Eq.~$\ref{virial}$) for runs of different resolutions. We check the deviation from zero of the quantity
\begin{equation}
\vert VC \vert=\frac{\vert 3\Pi+W+\mathscr{M}\vert}{\vert W\vert}.
\end{equation}
 Fig.~$\ref{VC}$ shows that $\vert VC \vert$ decreases with increase in the number of grid points whether or not the effect of Landau quantization is included. For the latter the code converges very well with $\vert VC \vert$ well below $10^{-5}$. When Landau quantization is taken into account, the code converges with $\vert VC \vert$ below $10^{-4}$. Here we have neglected the effect of energy quantization when $\nu$ is greater than $\sim20$.

\section{Results} 
\label{results}

We now summarize the results obtained with our numerical procedure described above.

\subsection{Configurations without Landau quantization: increase of maximum mass}
\begin{figure}
\centering
\includegraphics[width=0.45\textwidth]{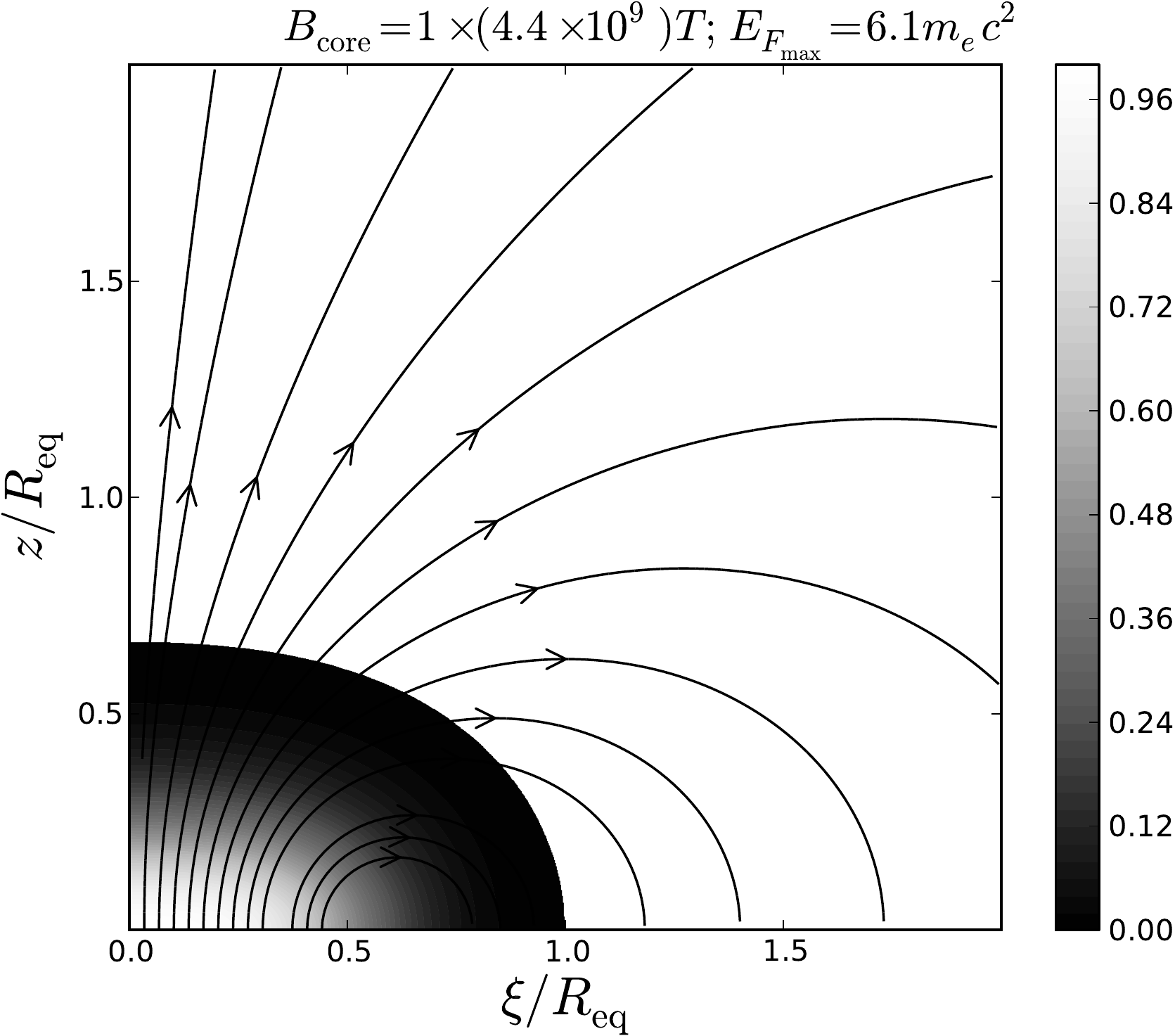}
\caption{Distribution of mass density (grey-scale) in units of central density and magnetic field lines (solid lines) in the meridional cross-section of a configuration with central mass density $\rho_{\rm max}=4.275\times10^{11}$~kg.m$^{-3}$ and core magnetic field strength $B_{\rm core}=4.414\times10^9$~T.  $R_{\rm eq}$ is the equatorial radius; $\xi$ and $z$ are distances from and along the axis of azimuthal symmetry. The configuration is non-spherical with poloidal magnetic field lines.}
\label{fig:m_Blines}
\end{figure}

\begin{figure}
\centering
\includegraphics[width=0.45\textwidth]{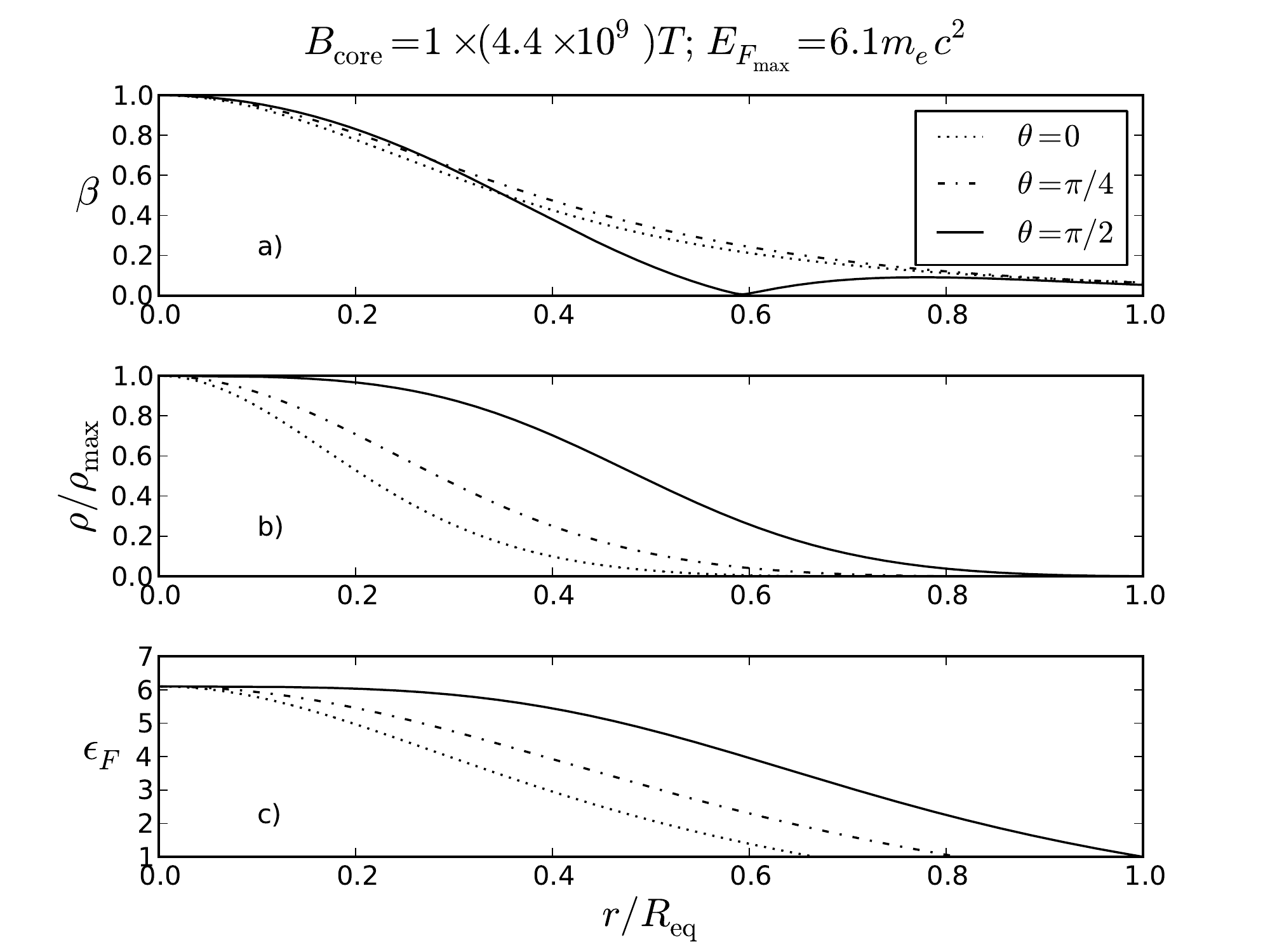}
\caption{The radial variation of a) magnetic field ($\beta=B/B_c$), mass density ($\rho/\rho_{\rm max}$) and c) Fermi energy ($\epsilon_{\rm F}$) along the axis of symmetry ($\theta=0$), at  $45^{\circ}$ to the symmetry axis and in the equatorial plane ($\theta=\pi/2$) of the white dwarf shown in Fig.~$\ref{fig:m_Blines}$.}
\label{B_rho_Hef6_1}
\end{figure}

If we ignore the effect of Landau quantization on the degeneracy pressure, then the mass distribution and the magnetic field configuration can be obtained for a specified core density and core magnetic field just using $\mathcal M=\mathcal M_0$. The stellar structure no longer remains spherical for high magnetic field as shown in Fig.~$\ref{fig:m_Blines}$. As expected, the magnetic field lines show poloidal characteristic and the field strength plot of Fig.~$\ref{B_rho_Hef6_1}$ shows a kink at the equatorial plane indicating a reversal of field direction. The mass distribution is not spherically symmetric due to the effect of Lorentz force in the presence of magnetic field. A comparison between radial distribution of magnetic field ($\beta=B/B_c$) and Fermi energy ($\epsilon_{\rm F}=E_{\rm F}/(m_{\rm e}c^2)$), shown in Fig.~$\ref{B_rho_Hef6_1}$, indicates that the radial gradient of $\epsilon_{\rm F}$ is higher than the radial gradient of $\beta$.

\begin{figure}
\centering
\includegraphics[width=0.45\textwidth]{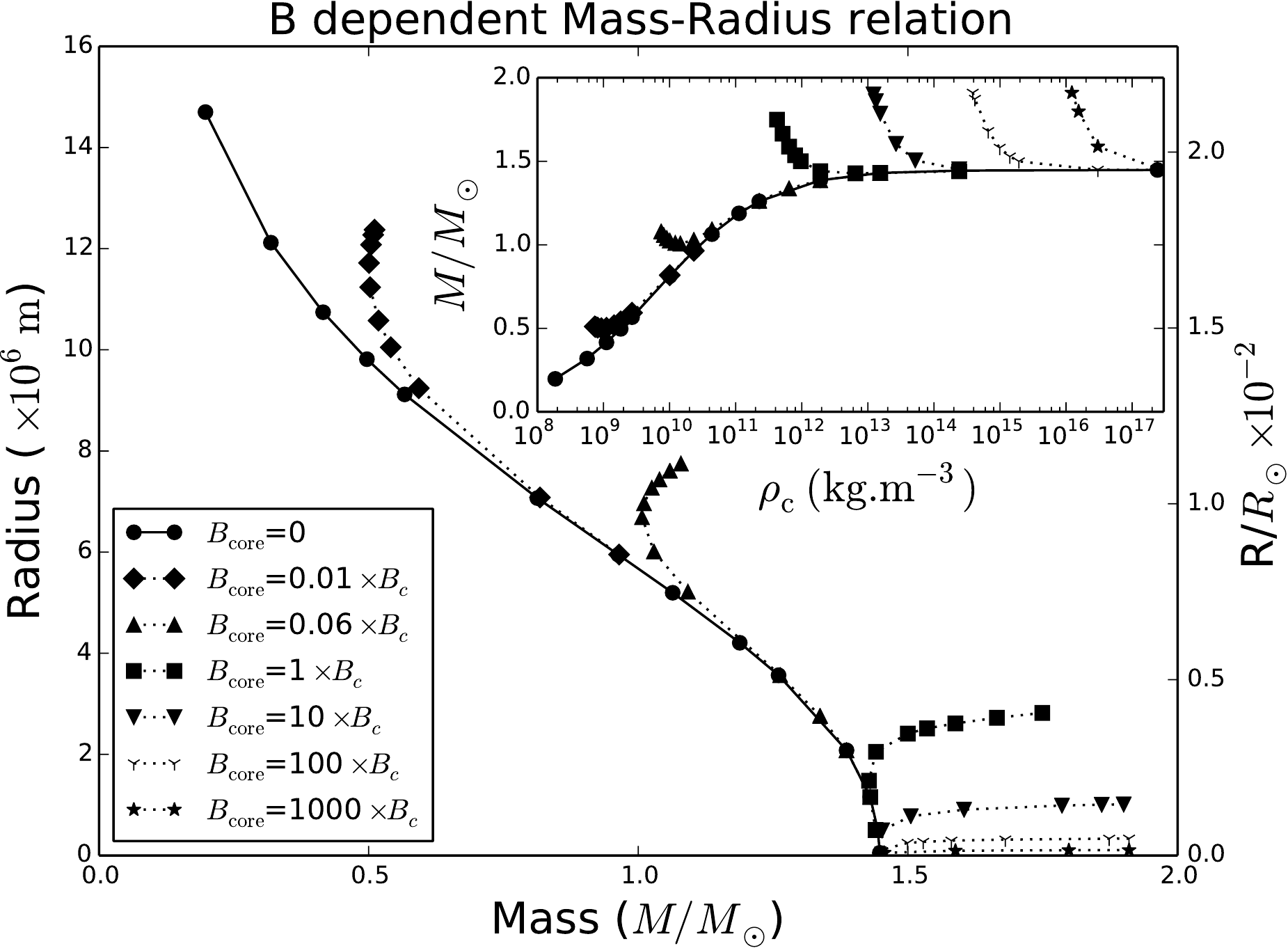}
\caption{Mass-radius relation of white dwarfs for different values of core magnetic field, with varying central density ($\rho_c$). The inset shows the dependence of mass on central density of these configurations. For a fixed core magnetic field, configurations with lower central density  deviate more strongly from the mass-radius relation of non-magnetic configurations.  The additional matter supported by magnetic forces can raise the mass limit to $\sim 1.9$~$M_{\odot}$.}
\label{M-R}
\end{figure}

We then construct a sequence of white dwarf configurations with a fixed core magnetic field but varying central density. From Fig.~$\ref{M-R}$ it is observed that as the central density is reduced the white dwarf mass decreases and radius increases in the absence of a strong magnetic field. For a fixed strong core magnetic field, the behaviour of mass and radius is similar to that of a non-magnetic configuration when the density is high. But as the central density decreases keeping the core field strength constant, equilibria at higher masses than in the absence of magnetic field can be obtained. It is also observed that below a certain central density there is no equilibrium configuration for a given core magnetic field.

The ability of a magnetic white dwarf to support more mass can be explained using the Lorentz term in hydrostatic equilibrium Eq.~($\ref{hydro_eui}$). In the integral form as expressed in Eq.~($\ref{intro_c}$), the magnetic field dependent function $\mathcal M$ modifies the Fermi energy, hence the degeneracy pressure, the mass density and consequently the gravitational potential. In our choice of poloidal magnetic field the Lorentz force $\left(\mathbf{f_m}=\mathbf{j}\boldsymbol{\times}\mathbf{B}\right)$ is zero along the symmetry axis and strong along the equatorial direction, with oppositely directed gravitational force $\left(\mathbf{f_g} =-\rho\boldsymbol\nabla\Phi\right)$ in the high density region. As a result, at the core of a strongly magnetized configuration, the Fermi energy gradient (and hence the mass density variation) along the equatorial direction  is much smaller than that along the axis (Fig.~$\ref{B_rho_Hef6_1}$).

\begin{figure}
\centering
\includegraphics[width=0.45\textwidth]{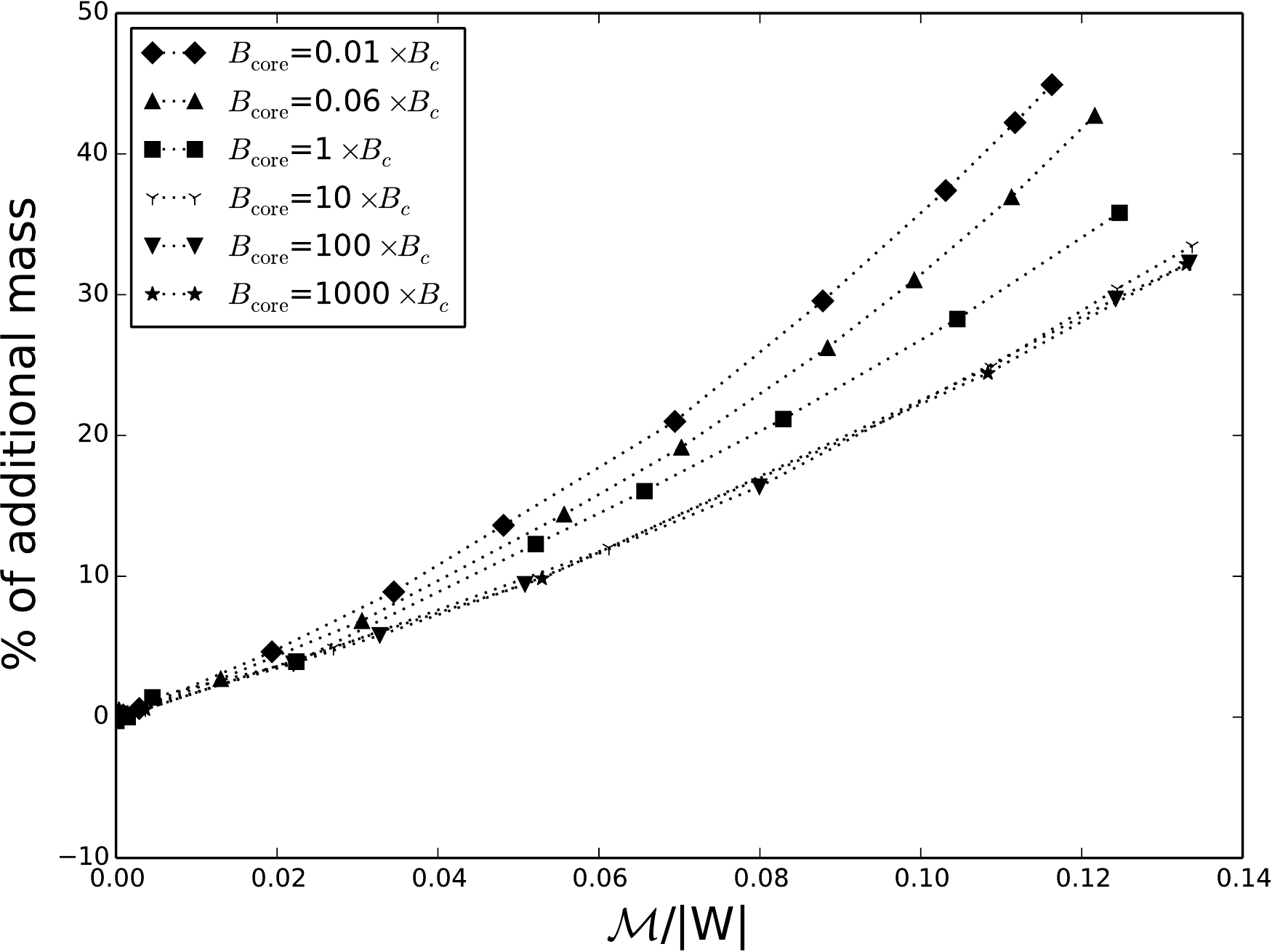}
\caption{The fraction of supported additional mass ($\Delta M/M$) in per cent plotted against the ratio of magnetic energy to gravitational energy of configurations with different core magnetic field strengths. Configurations with ultra-strong core fields are highly relativistic, and in this limit  $\Delta M/M$ is nearly independent of the magnetic field strength.}
\label{ExMass_Emag}
\end{figure}

We define the additional mass ($\Delta M$) supported by the magnetic field to be the mass difference of configurations of the same central density with and without magnetic field. To illustrate the dependence of $\Delta M$ on the field strength, we plot the dimensionless quantity $\Delta M/M$ against $\mathscr{M}/|W|$ in Fig.~$\ref{ExMass_Emag}$. It is observed that the extra mass supported in the presence of magnetic fields is closely related to the ratio of total magnetic energy to gravitational energy of the configuration. The excess mass depends, in general, on both $B_{\rm core}$ and $\mathscr{M}/|W|$, but for highly relativistic configurations the dependence on $B_{\rm core}$ nearly disappears.  The end points of the sequences with different core field strengths indicate that strongly relativistic configurations with magnetic energy around $12-13$ per cent of the gravitational energy can support more than 0.4$M_{\odot}$ over non-magnetic configurations extending the maximum mass to $\sim1.9M_{\odot}$.

\begin{figure*}
\centering
\includegraphics[width=0.95\textwidth]{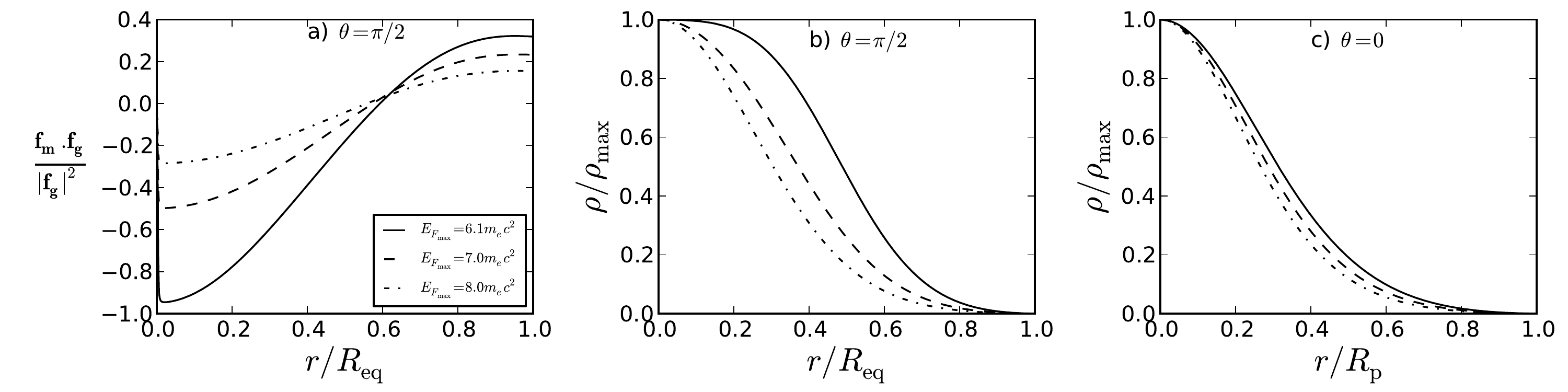}
\caption{Comparison between configurations with different central Fermi energy (equivalent to central density) but the same core magnetic field strength $B_{\rm core}=4.414\times10^9$~T. a) Relative strength of the Lorentz force component along the gravitational force, evaluated in the equatorial plane. Negative component ratio indicates oppositely directed force components. b) Radial profile of mass density in the equatorial plane c) Density profile along the symmetry axis. Density profile in the equatorial plane is more strongly modified due to the presence of Lorentz force.}
\label{fm_fg}
\end{figure*}

The gravitational force within a bound object is directed towards the centre of mass of the system. Given our choice of current density (along $\boldsymbol{\hat\phi}$), the resultant Lorentz force is oriented in the $\xi-z$ plane. The outward radial component of the Lorentz force would unbind the star if it exceeds the gravitational force at any point of the configuration. This leads to a lower bound on the central density for a given field strength, as the lowering of the density decreases the gravitational force contributing to such a loss of equilibrium. This is clearly seen in Fig.~$\ref{fm_fg}$a. Fig.~$\ref{fm_fg}$b shows that the mass distribution within the configuration is strongly modified in the equatorial plane in the presence of Lorentz force, while on the axis of symmetry the mass distribution is hardly affected (Fig.~$\ref{fm_fg}$c). An equilibrium solution can be obtained when the Lorentz force component does not exceed the gravitational force. Our numerical technique allowed us to continue the equilibrium computations up to a stage where the Lorentz force component reached nearly $98$~per cent of the gravitational force at some point within the star. Proceeding beyond this point proved challenging, but an extrapolation of our sequence of results to the full equality of these two force components suggests that the mass of the extreme configurations computed by us (Fig.~$\ref{M-R}$) would be within 1 per cent of the true maximum value.   

\subsection{Effect of Landau quantization on white dwarf structure}

\begin{figure}
\centering
\includegraphics[width=0.45\textwidth]{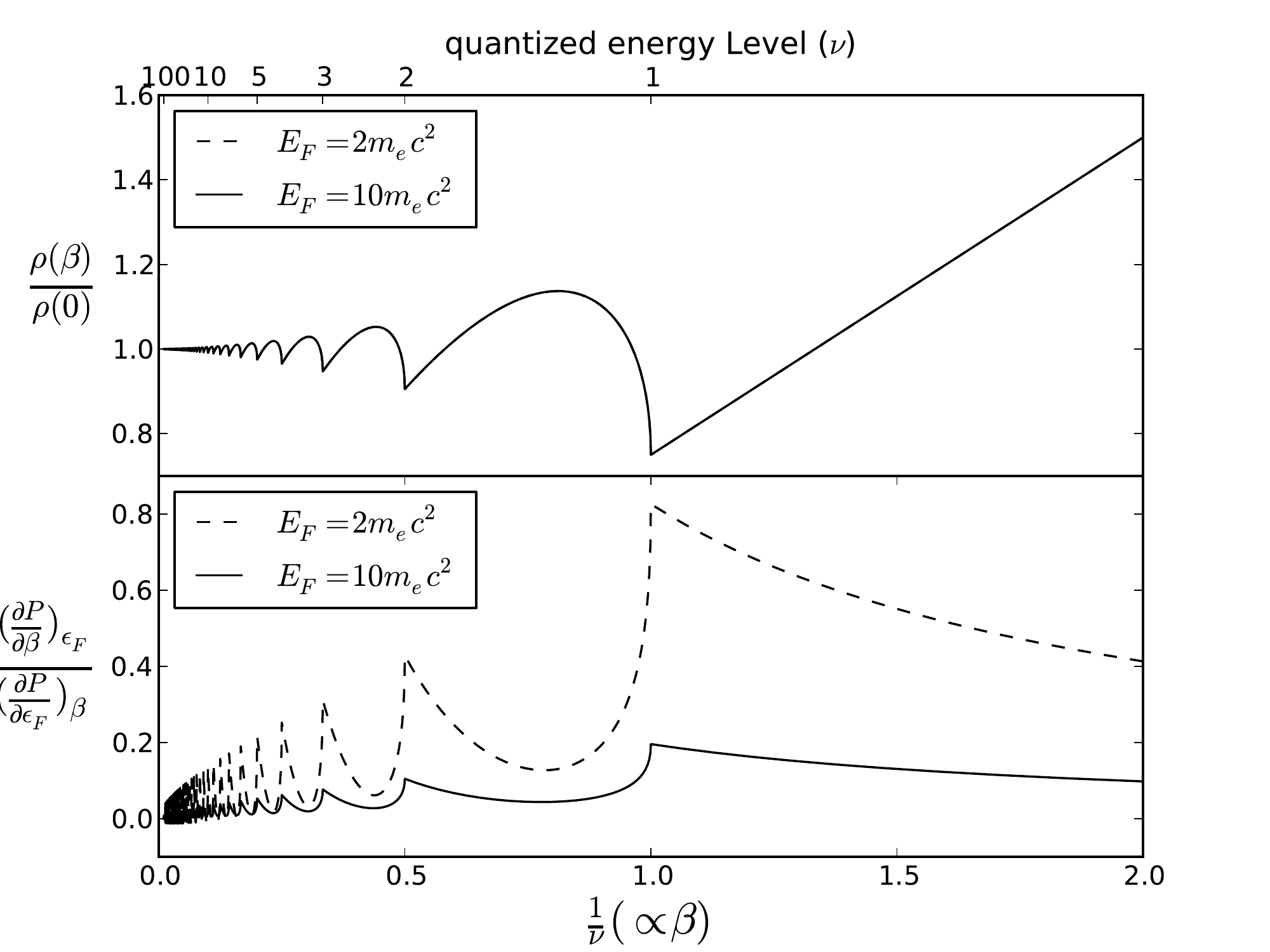}
\caption{Effect of Landau quantization on the equation of state.  Upper panel: Magnetic field dependence of the mass density in presence of Landau quantization. The horizontal axis represents $\frac{1}{\nu}$ which is proportional to magnetic field strength $\beta\left(=\frac{\epsilon_{\rm F}^2-1}{2\nu}\right)$.  Labels on the top represent $\nu$, the highest energy level that would be accessed for the corresponding magnetic field and Fermi energy. The vertical axis is density normalized to the non-magnetic case. Lower panel: relative  dependence of degeneracy pressure on magnetic field and mass density for an equation of state with Landau quantization.  Curves for two different values of Fermi energy are shown.  In the upper panel these two curves overlap completely.}

\label{rhoP_B}
\end{figure}

The inclusion of Landau quantization in the equation of state modifies the distribution of mass density and the electron degeneracy pressure within the star. Fig.~$\ref{rhoP_B}$ shows that in regions where high energy levels can be accessed, this modification is of insignificant magnitude.  However when the number of levels accessed is small, the mass density shows a strong dependence on the magnetic field.

Magnetic field dependent degeneracy pressure influences the stellar structure through the pressure gradient ($\frac{\partial P}{\partial \beta}\boldsymbol\nabla\beta$). This term dominates over $\frac{\partial P}{\partial \epsilon_F}\boldsymbol\nabla\epsilon_{\rm F}$ when the Fermi energy is low ($\epsilon_F \rightarrow 1$) and the magnetic field is strong. It is observed from Fig.~$\ref{rhoP_B}$ and Fig.~$\ref{B_rho_Hef6_1}$ that this situation arises only near the stellar surface, where electrons are confined to the lowest Landau levels. For a fixed Fermi energy, as the magnetic field strength ($\beta=\frac{\epsilon_F^2-1}{2} \frac{1}{\nu}$) decreases, the factor $\frac{\partial P}{\partial \beta}$ approaches zero. Electrons in the core occupy at least $\nu=$17-20 even for the structures  strongly dominated by magnetic field, namely those near the endpoint of $M-R$ relations displayed in Fig.~$\ref{M-R}$. As a result the overall effect of Landau quantization on the white dwarf structure is in fact minimal. We quantify this below.

\begin{figure}
\centering
\includegraphics[width=0.45\textwidth]{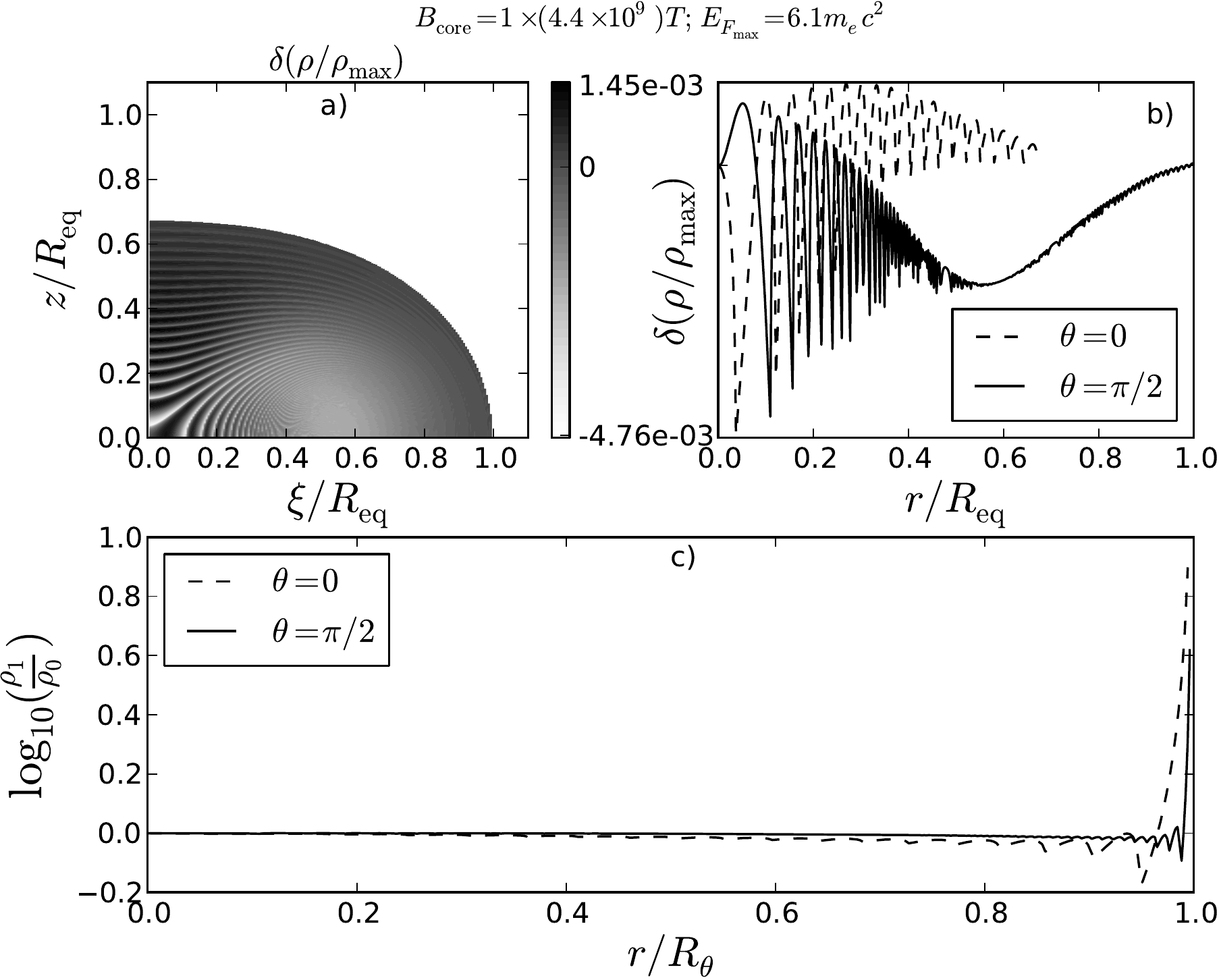}
\caption{ Relative modification of mass distribution of the configuration of Fig.~$\ref{fig:m_Blines}$ due to the inclusion of Landau quantization in the equation of state. Here $\delta(\rho/\rho_{\rm max})=\rho_1/\rho_{\rm max}-\rho_0/\rho_{\rm max}$, $\rho_0$ and $\rho_1$ being the mass density of the configurations without and with the inclusion of Landau quantization and $\rho_{\rm max}$ is the maximum density of either configuration. a) Distribution of $\delta(\rho/\rho_{\rm max})$ through the star and b) Equatorial (solid) and polar (dashed) cuts through the distribution shown in panel a). 
The relative mass density differs less than 1 per cent between the quantised and the non-quantised configurations. The plot of $\rho_1/\rho_0$ in panel c) shows that this difference is significant only very near the stellar surface.  Here, $R_{\theta}$ is the distance of the stellar boundary from the centre at an angle $\theta$ from the symmetry axis.}
\label{drho_dbP_B1}
\end{figure}

To obtain the stellar structure when the degeneracy pressure is magnetic field dependent due to Landau quantization, we use $\mathcal M=\mathcal M_0+\mathcal M_1$ in Eq.~($\ref{intro_c}$). The configurations obtained for the same core magnetic field strength and central Fermi energy are almost identical to those without Landau quantization. As shown in Fig.~$\ref{drho_dbP_B1}$, the relative difference in their mass density is appreciable only at a thin layer near the stellar surface. However, the density itself is very low in this layer and hence this contributes very little to the total mass. The mass of the configuration, with or without the effect of Landau quantization, remains the same to within less than 0.1 per cent.

\begin{table*}
\begin{tabular}{c|ccccccc}
\hline
\bf central conditions &Landau quantization & \bf $\rho_c$&  $M/M_\odot$ & \bf $R_{\rm eq}/10^6m$ & \bf $R_{\rm p}/R_{\rm eq}$ & \bf $\mid\mathscr{M}/W\mid$ & \bf $\mid VC\mid$\\ \hline

  $E_{\rm{F}{\rm max}}$=6.1$m_ec^2$ &not considered &145.2618 &1.7496 & 2.8020 &0.665  &0.1247 &$5.4088\times10^{-6}$\\ \cline{2-8}
  $B_{\rm core}=4.414\times10^9~T$ & considered &145.3913 &1.7506 & 2.8057 &0.673 &0.1245 &$3.1047\times10^{-5}$\\ \hline
 
  $E_{\rm{F}{\rm max}}$=59$m_ec^2$ & not considered &136860.3 &1.8995 &0.3338 &0.680 &0.1295 &$9.3629\times10^{-06}$\\ \cline{2-8} 
  $B_{\rm core}=4.414\times10^{11}~T$& considered &137128.2 &1.9008 & 0.3411 &0.705 &0.1289 &$1.9038\times10^{-05}$\\ \hline

\end{tabular}
\caption{Comparison of configurations with and without the consideration of Landau quantization (grid=512$\times$512). Here  $\rho_{\rm max}= \rho_c\times2.9\times10^9$~kg.m$^{-3} $ is the maximum mass density, $M$ is the total mass of the configuration, $R_{\rm eq}$ and $R_{\rm p}$ are the equatorial and polar radii and $\mathscr{M}$ and  $W$ are the total magnetic and gravitational energy respectively.}\label{conf_comp}
\end{table*}

Table~$\ref{conf_comp}$ shows that the physical parameters of the maximal configurations change very little with the inclusion of Landau quantization.  For the two configurations listed, namely with $\beta_c$ = 1 and 100, Landau quantization affects the central density by approximately 0.1-0.2 per cent and $\mathscr{M}/|W|$ ratio by less than 0.5 per cent as the mass and the radius remain nearly the same. One noticeable effect of Landau quantization is the reduction of asphericity, as the ratio of $R_{\rm p}$ to $R_{\rm eq}$ is seen to increase by up to several percent.

\subsection{The form of the current distribution}
The functions $f(u)$ and $\mathcal M(u)$ of Eq.~($\ref{mu_0j_}$) are arbitrary functions of $u$. So far we have used the functional form as mentioned in Eq.~($\ref{func_fu}$) and Eq.~($\ref{func_Mu}$). However, as discussed by \cite{fujis12}, these functions may be chosen to have a more general form as follows:
\begin{align}
 \text{$\mathcal M(u)$}&=\frac{\alpha_0}{1+m}(u+\epsilon)^{m+1} \label{funcg_Mu}\\
 f(u)&=
 \begin{cases}
  \frac{\kappa_0}{k+1}(u-u_{\rm max})^{k+1} & \text{if } u \geq u_{\rm max}, \\
  0, & \text{if } u \leq u_{\rm max}.
 \end{cases}
\end{align}
here $\alpha_0$, $\kappa_0$, $m$, $k$ and $\epsilon$ are constants, $u_{\rm max}$ is the maximum value of $u$ on the stellar surface, $\epsilon=1\times10^{-6}$ is used to avoid the divergence of the function $\mathcal M$.

A non-zero value of $f(u)$ within stellar interior gives the toroidal component of the magnetic field. Our computations with different forms of $f(u)$ suggest that such configurations can support at most a few percent of the total magnetic field energy in the toroidal component, as also found by \cite{fujis12}. The toroidal fraction becomes smaller with the rise of the poloidal strength. Thus the inclusion of the toroidal component does not significantly affect the estimate of maximum field strength in static equilibrium.

The parameter $m$ controls the functional form of $\mathcal M(u)$.  For $m$=0 (and $\epsilon\rightarrow 0$) Eq.~($\ref{funcg_Mu}$) reduces to the form used in Eq.~($\ref{func_Mu}$). The poloidal magnetic field strength rises sharply towards the centre for $m<$-1, and becomes more uniform with increasing $m$  \citep{fujis12}. For the $m$=0 configuration studied here, $B$ is maximum at the core and the current density ($j$) peaks at a radius in between the centre and the surface. This causes the Lorentz force density ($\mathbf{f_m}$) to peak at an equatorial radius in between the centre and the location of  maximum $j$. The position of the Lorentz force peak moves outwards with increasing $m$. The gravitational force density ($\mathbf{f_g}$) also has a peak in between the centre and the surface, as it is zero at both these limits.  It turns out that the peak locations of the Lorentz force density and the gravitational force density are well matched for $m\sim 0$, making this a case well suited to the study of configurations with maximal magnetic support.

\section{Discussion and Conclusions} 
\label{conclusion}

In this paper we have computed the structure of stationary, axisymmetric, perfectly conducting electron degenerate white dwarfs with strong poloidal magnetic field, and have presented the resulting mass-radius relations. Our findings are:
\renewcommand{\theenumi}{\roman{enumi})}
\begin{enumerate}
\item The white dwarf of a given central density can support a larger mass in the presence of a strong magnetic field as the Lorentz force can augment the pressure gradient balancing gravity. Solutions for static configurations can be obtained as long as the local Lorentz force density does not exceed the gravitational force density anywhere within the star.
\item For a fixed core magnetic field, as the central density is decreased from its maximal value, the mass and the radius of a white dwarf initially follow a relation similar to that of a non-magnetic configuration. However the relation starts to depart strongly from the non-magnetic case as the gravitational energy of the configuration drops to values comparable to the magnetic energy, since additional mass can then be supported by the Lorentz force.
For a purely poloidal field structure a strongly relativistic white dwarf can support an additional mass of up to $\sim 0.5M_{\odot}$, when the magnetic field energy of the configuration is approximately $13$ per cent of the gravitational energy. 
\item Even at the maximum strength of the magnetic field, the impact of Landau quantization on the stellar structure is found to be not significant; the physical parameters of configurations computed with and without the inclusion of Landau quantization differ by less than one per cent.
\end{enumerate}

It is to be noted that the results presented above relate only to magnetostatic equilibrium conditions. The stability of these equilibria are not addressed here. At extremely high matter density neutronization due to electron capture may ensue, reducing the electron degeneracy pressure support. The threshold Fermi energy for neutronization for Helium, Carbon and Oxygen correspond to critical densities in the range $10^{13-14}$ kg.m$^{-3}$ \citep{shapi83,chame13,chamel_2014}. We find that configurations with $B_{\rm core}=10-20~B_c$ show maximum mass close to 1.9 $M_{\odot}$ (Fig.~$\ref{M-R}$) even for densities below this limit. However magnetically supported configurations such as these may be prone to several dynamical instabilities.  For example, if the interior field distribution is not force-free, then reorganisation of the field structure may proceed in Alfv{\'e}n time scale if the interior behaves as a perfect fluid, or in viscous/elastic time scale if that happens to be longer.  Dissipation of currents due to ohmic effects, assisted by Hall cascade in favourable circumstances \citep{gold_reis92}, would affect the strength and distribution of the interior magnetic field over the long term.  For typical physical parameters of the strongly magnetized configurations considered here, the Alfv{\'e}n time evaluates to $\sim 0.1$~s, the viscous time to $\sim 10^{17}$~s and the ohmic time to $\sim 10^{18}$~s (using expressions for viscosity from \citep{duris73} and resistivity from \cite{yakov80} and \cite{cummi02}).  Interchange instabilities such as Parker modes may also become important for specific distributions of magnetic field and matter. \cite{braithwaite09} and \cite{braithwaite06} demonstrate that for a low field axisymmetric stable polytropic star the upper limit on the energy of the poloidal field is about 80 per cent of total magnetic energy. Further investigation of this for the configurations presented here will be undertaken in a forthcoming work. Nevertheless, It does appear that at least short-lived super-Chandrasekhar mass white dwarfs may indeed form via magnetic support and possibly contribute to over-luminous SNIa.

\section{Acknowledgement}
We thank Dr. Nicolas Chamel and the referee, Dr. Christopher Tout, for their valuable comments that helped us to improve the paper. PB thanks Sushan Konar for introducing this problem and CSIR, India for a Junior Research Fellowship.


\def\aj{AJ}%
\def\actaa{Acta Astron.}%
\def\araa{ARA\&A}%
\def\apj{ApJ}%
\def\apjl{ApJ}%
\def\apjs{ApJS}%
\def\ao{Appl.~Opt.}%
\def\apss{Ap\&SS}%
\def\aap{A\&A}%
\def\aapr{A\&A~Rev.}%
\def\aaps{A\&AS}%
\def\azh{AZh}%
\def\baas{BAAS}%
\def\bac{Bull. astr. Inst. Czechosl.}%
\def\caa{Chinese Astron. Astrophys.}%
\def\cjaa{Chinese J. Astron. Astrophys.}%
\def\icarus{Icarus}%
\def\jcap{J. Cosmology Astropart. Phys.}%
\def\jrasc{JRASC}%
\def\mnras{MNRAS}%
\def\memras{MmRAS}%
\def\na{New A}%
\def\nar{New A Rev.}%
\def\pasa{PASA}%
\def\pra{Phys.~Rev.~A}%
\def\prb{Phys.~Rev.~B}%
\def\prc{Phys.~Rev.~C}%
\def\prd{Phys.~Rev.~D}%
\def\pre{Phys.~Rev.~E}%
\def\prl{Phys.~Rev.~Lett.}%
\def\pasp{PASP}%
\def\pasj{PASJ}%
\def\qjras{QJRAS}
\def\rmxaa{Rev. Mexicana Astron. Astrofis.}%
\def\skytel{S\&T}%
\def\solphys{Sol.~Phys.}%
\def\sovast{Soviet~Ast.}%
\def\ssr{Space~Sci.~Rev.}%
\def\zap{ZAp}%
\def\nat{Nature}%
\def\iaucirc{IAU~Circ.}%
\def\aplett{Astrophys.~Lett.}%
\def\apspr{Astrophys.~Space~Phys.~Res.}%
\def\bain{Bull.~Astron.~Inst.~Netherlands}%
\def\fcp{Fund.~Cosmic~Phys.}%
\def\gca{Geochim.~Cosmochim.~Acta}%
\def\grl{Geophys.~Res.~Lett.}%
\def\jcp{J.~Chem.~Phys.}%
\def\jgr{J.~Geophys.~Res.}%
\def\jqsrt{J.~Quant.~Spec.~Radiat.~Transf.}%
\def\memsai{Mem.~Soc.~Astron.~Italiana}%
\def\nphysa{Nucl.~Phys.~A}%
\def\physrep{Phys.~Rep.}%
\def\physscr{Phys.~Scr}%
\def\planss{Planet.~Space~Sci.}%
\def\procspie{Proc.~SPIED}%
\let\astap=\aap
\let\apjlett=\apjl
\let\apjsupp=\apjs
\let\applopt=\ao

\bibliographystyle{mn2e}	
\bibliography{ref_DB.bib}
\label{lastpage}
\end{document}